\documentclass[a4paper,11pt]{article}
\pdfoutput=1 

\usepackage{jheppub} 

\usepackage{jheppub} 

\usepackage{comment}
\usepackage{enumerate}
\usepackage{physics}
\usepackage{color}
\usepackage{lineno}

\title{Analytical estimation of the signal to noise ratio efficiency in axion dark matter searches using a Savitzky-Golay filter}
\author[a,b,1]{A. K. Yi,\note{Now at SLAC National Accelerator Laboratory, 2575 Sand Hill Rd., Menlo Park, California 94025, USA}}
\author[b]{S. Ahn,}
\author[b,2]{B. R. Ko,\note{Corresponding author.}}
\author[b,a]{and Y. K. Semertzidis\,}
\affiliation[a]{Department of Physics, Korea Advanced Institute of
	Science and Technology (KAIST), Daejeon 34141, Republic of Korea}
\affiliation[b]{Center for Axion and Precision Physics Research,
	Institute for Basic Science (IBS), Daejeon 34141, Republic of Korea}

\emailAdd{andrewyi@slac.stanford.edu}
\emailAdd{saebyeokahn@ibs.re.kr}
\emailAdd{brko@ibs.re.kr}
\emailAdd{yannis@ibs.re.kr}

\abstract{
  The signal to noise ratio efficiency $\epsilon_{\rm SNR}$ in axion
  dark matter searches has been estimated using large-statistic
  simulation data reflecting the background information and the expected
  axion signal power obtained from a real experiment. This usually
  requires a lot of computing time even with the assistance of
  powerful computing resources.
  Employing a Savitzky-Golay filter for background subtraction, in this
  work, we estimated a fully analytical $\epsilon_{\rm SNR}$ without
  relying on large-statistic simulation data, but only with an arbitrary
  axion mass and the relevant signal shape information.
  Hence, our work can provide $\epsilon_{\rm SNR}$ using minimal
  computing time and resources prior to the acquisition of
  experimental data, without the detailed information that has to be
  obtained from real experiments.  
  Axion haloscope searches have been observing the
  coincidence that the frequency independent scale factor $\xi$ is
  approximately consistent with the $\epsilon_{\rm SNR}$. This was
  confirmed analytically in this work, when the window length of the
  Savitzky-Golay filter is reasonably wide enough, i.e., at least 5
  times the signal window.  
}


\begin{document}
	
\maketitle
\flushbottom
	
\section{Introduction}
According to cosmological measurements and the standard model of Big
Bang cosmology~\cite{PLANCK}, cold dark matter (CDM) is responsible
for about 85\% of the total matter in our Universe. CDM does
not belong to the Standard Model of particle physics (SM), and
as of today the nature of CDM is unknown in spite of the
strong evidence of its existence~\cite{CDM-EVIDENCE}.        
The axion~\cite{AXION} is one of the iconic CDM candidates and
originally stems from a breakdown of a new symmetry proposed by Peccei
and Quinn~\cite{PQ}, as a very natural solution to the
strong $CP$ problem in the SM~\cite{strongCP}.        
In light of the current galaxy formation dark matter has to be
massive, stable, and nonrelativistic, and the axion is
predicted to meet all of these conditions.        
The axion haloscope search proposed by Sikivie utilizes
axion-photon coupling and a microwave resonant cavity, which results
in the resonant conversion of axions to photons~\cite{sikivie}. This
enhances the detected axion signal power drastically. Making the
axion haloscope the most promising method for axion dark matter
searches in the microwave region.
Recent experimental efforts have reached the
Dine-Fischler-Srednicki-Zhitnitskii (DFSZ) axion
sensitivity~\cite{ADMX-DFSZ, 12TB-PRL}, where the DFSZ axion can be
implemented in grand unified theories (GUT)~\cite{GUT}. If DFSZ cold
axion dark matter turns out to constitute 100\% of the local dark
matter density, that would not only explain the total matter in our
Universe, but also support GUT.

Since the resonated axion signal power is only sensitive to the resonant
frequency region, and information about axion mass is absent,
the most significant and practical figure of merit in axion haloscope
searches is the scanning rate~\cite{scanrate}. One of the experimental
parameters determining the scanning rate is the signal to noise ratio
(SNR) efficiency squared $\epsilon^2_{\rm SNR}$~\cite{JINST}, where
the ${\rm SNR}=P^{a\gamma\gamma}_a/\sigma_{P_n}$
according to the radiometer equation~\cite{DICKE} and we define
$\epsilon_{\rm SNR}$ later in equation~(\ref{eq_snr_eff}).
In the SNR equation,
$P^{a\gamma\gamma}_a$ is the expected axion signal power for an
axion-photon coupling strength~\cite{sikivie, scanrate} and
$\sigma_{P_n}$ is the fluctuation in the noise power $P_n$.
Estimates of the $\epsilon_{\rm SNR}$ have relied on
large-statistic simulation data that usually require a lot of
computation time and storage resources~\cite{ADMX, HAYSTAC, JHEP}.
With hints reported in refs.~\cite{HAYSTAC, JHEP}, in this study we
have developed an analytical method to estimate the
$\epsilon_{\rm SNR}$ for axion dark matter searches.
The $\epsilon_{\rm SNR}$ for axion haloscope searches is
frequency-independent or, equivalently, background-independent with
the relevant scale factor $\xi$, as demonstrated by
refs.~\cite{HAYSTAC, JHEP},
where $\xi$ is necessary and affected by the combination of power
excess bins that are correlated with each other due to background
subtraction. In addition, one of our previous works showed that
$\epsilon_{\rm SNR}$ is practically the signal power efficiency due to
background subtraction~\cite{JHEP}.
In another one of our recent works~\cite{12TB-PRD}, however, we found
a subtle effect to the background fluctuation due to background
subtraction with a specific condition, which was considered as well.

Inspired by that information, we calculated the signal power
efficiency $\epsilon_{P^{a\gamma\gamma}_a}=\epsilon_{\rm sig}$ whose
effects can first be seen in Fig.~\ref{fig_signal_before_coadd} and
$\xi$ independently using only an arbitrary axion mass and the
relevant signal shape information.
They showed good agreement with those obtained from
large-statistic simulation data.        
In the end, axion haloscope searches have been observing the
coincidence that the frequency independent scale factor $\xi$ is
approximately consistent with the $\epsilon_{\rm SNR}$. This was
confirmed analytically in this work, if the window length of our
background estimator is reasonably wide enough, i.e., at least 5
times the signal window.

\section{Obtaining the SNR efficiency analytically}
The general data analysis procedure for axion haloscope searches
studied in this work aims to maximize the significance of an axion
signal by weighting and combining data points~\cite{HAYSTAC}. Raw data
is acquired through power spectra in the frequency domain. The
digitized power spectrum data will have the background noise power
stored in frequency bins that are spaced at intervals of the resolution
bandwidth (RBW) $\Delta\nu$. The background will feature a baseline
characterized by the system noise and its fluctuations. If an axion
signal exists amidst the background, it will appear as a lineshape
which is distributed to multiple frequency bins. In the case of the
standard halo model, the lineshape is a boosted
Maxwellian~\cite{TURNER}
\begin{equation} \label{EQ:LINESHAPE}
	f(\nu) = \frac{2}{\sqrt{\pi}} \left( \sqrt{\frac{3}{2}} \frac{1}{r} \frac{1}{\nu_{a}\expval{\beta^{2}}} \right) \sinh \left( 3r \sqrt{ \frac{2(\nu - \nu_{a})}{\nu_{a} \expval{\beta^{2}}} } \right) \exp \left( -\frac{3(\nu - \nu_{a})}{\nu_{a}\expval{\beta^{2}}} - \frac{3r^{2}}{2}\right).
\end{equation}

This lineshape depends on the axion frequency $\nu_{a}$, $r = v_{E}/v_{\mathrm{rms}}$,
and $\expval{\beta^{2}} = 1.5v_{\mathrm{rms}}^{2}/c^{2}$, where
$v_{E}$ is the velocity of Earth with respect to the galaxy halos,
$v_{\mathrm{rms}}$ is the local circular velocity of the galaxy, and
$c$ is the speed of light. In the analysis stage an axion signal
window $\Delta\nu_a$ is used where the integral of $f(\nu)$ from $\nu_{a}$
to $\nu_{a} + \Delta\nu_{a}$ nears unity. The axion signal power is
weak; for instance in the CAPP-12TB experiment it was on the
order of tens of yoctowatts~\cite{12TB-PRL}, making it difficult to
distinguish from noise fluctuations. Therefore the data is further
processed in a way that increases the SNR of an expected signal.

\begin{figure}[h]
	\centering
	\includegraphics[width=0.48\columnwidth]{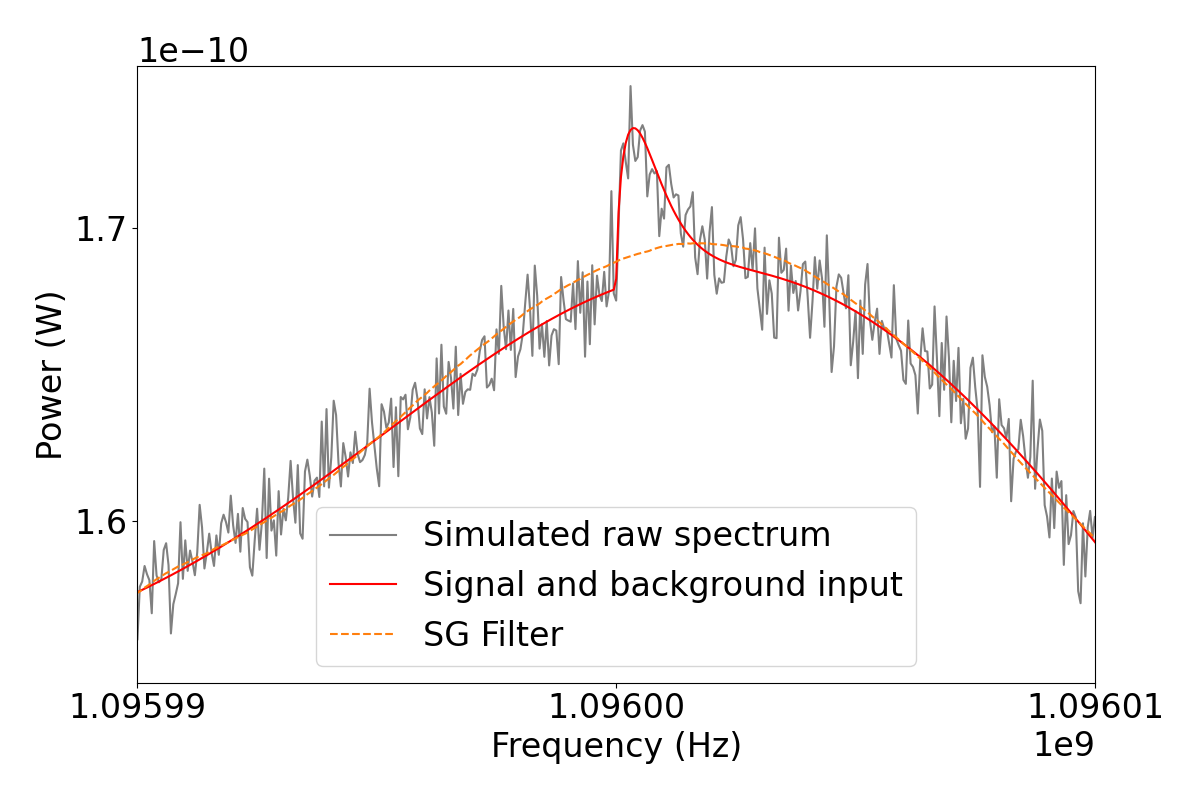} \quad
	\includegraphics[width=0.48\columnwidth]{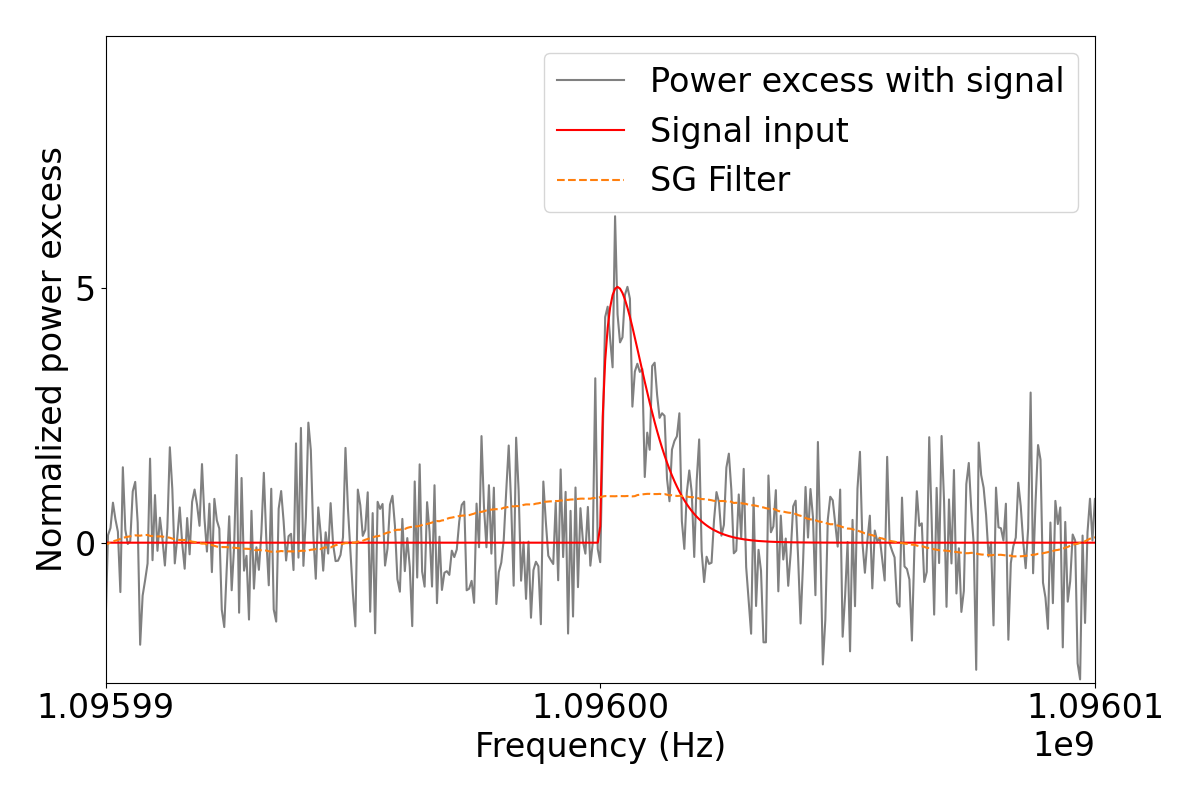}
	\caption
	{The effect of an axion signal distorting a background
          estimation using an SG filter (left) and its effect in terms
          of normalized power excess (right). An axion signal with
          exaggerated signal power was added at 1.096 GHz.}        
	\label{fig_signal_before_coadd}
\end{figure}

Following the conventional methods of haloscope data analysis, an
axion signal, should it exist, will appear as a sharp peak on the
background as shown in Fig.~\ref{fig_signal_before_coadd}.
This is also replicable by inserting a software-injected axion signal
into an experiment's background. Then the SNR of an input signal,
which will have its background perfectly removed, can be compared with
the SNR of signal which has its background removed by a fit. Examples
of a fit estimator are a $\chi^{2}$ fit such as in
refs.~\cite{12TB-PRL, CAPP-8TB-PRL} or a Savitzky-Golay (SG)
filter~\cite{SGFILTER} as seen in refs.~\cite{12TB-PRD, HAYSTAC}. In
this work we will be focusing on the latter as it does not require a
functional form of the background structure, can handle
zero-fluctuation data like the axion signal power considered in axion
dark matter searches, and its estimates are fully predictable.

Typically any estimate of the background will be affected by a signal
such as the one shown in Fig.~\ref{fig_signal_before_coadd} and due to
this the output SNR will not recover 100\% of the input. The
difference in the input and output SNR can be expressed through an
efficiency term $\epsilon_{\rm SNR}$
\begin{equation} \label{eq_snr_eff}
	\epsilon_{\rm SNR} = \frac{\mathrm{SNR}_{\mathrm{output}}}{\mathrm{SNR}_{\mathrm{input}}}\frac{1}{\xi} \simeq \frac{\epsilon_{\rm  sig}}{\xi} ,
\end{equation}
where we define $\epsilon_{\rm  sig} = \delta_{\rm output}/\delta_{\rm input}$
for input and output signal powers $\delta_{\rm input}$ and
$\delta_{\rm output}$. $\epsilon_{\rm  sig}$ is approximately equal to
$\mathrm{SNR}_{\mathrm{output}} / \mathrm{SNR}_{\mathrm{input}}$ as
$\sigma_{P_{n}}$ for input and output are practically
equal~\cite{JHEP}. The frequency-independent scale factor $\xi$ is
used to offset the bin-to-bin correlations that are introduced from
background subtraction~\cite{HAYSTAC, JHEP}. This value is set to the
width of the Gaussian fit for the normalized power excess of the
output. In terms of input, the normalized power excess' Gaussian width
should be equal to one.

\begin{figure}[h]
	\centering
	\includegraphics[width=0.48\columnwidth]{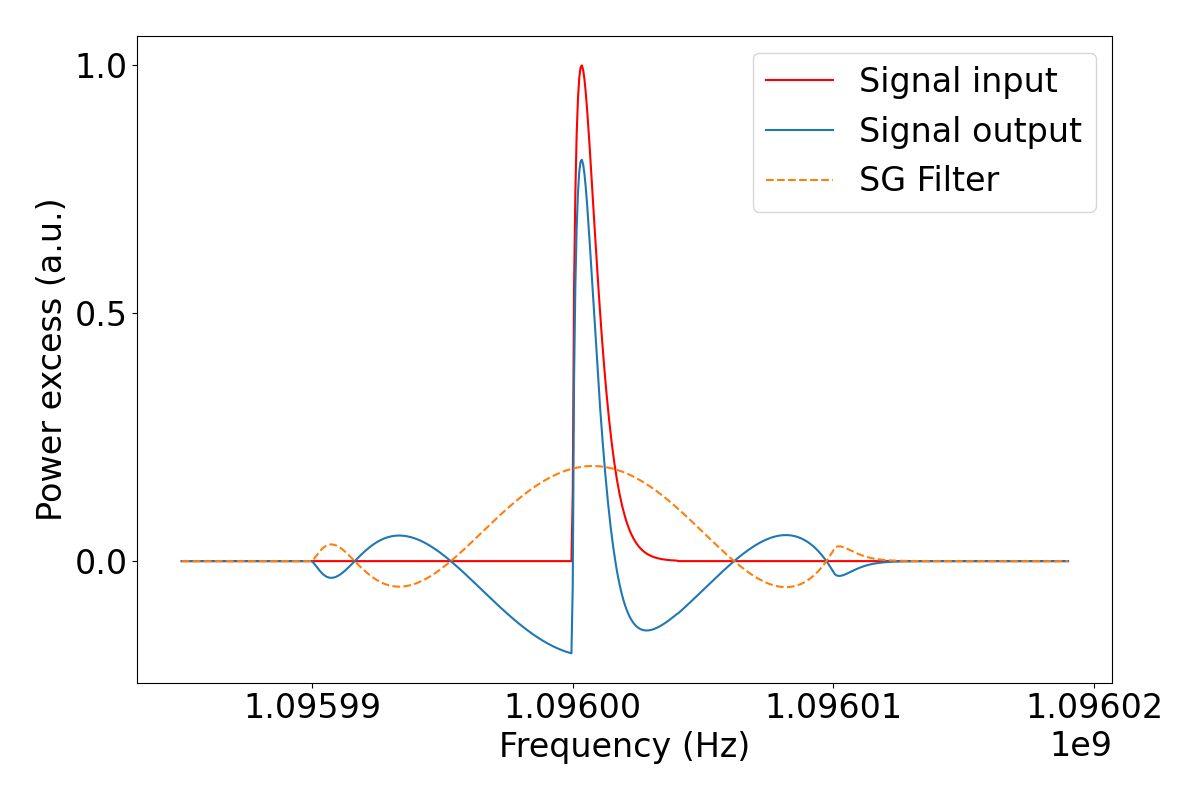}  \quad
	\includegraphics[width=0.48\columnwidth]{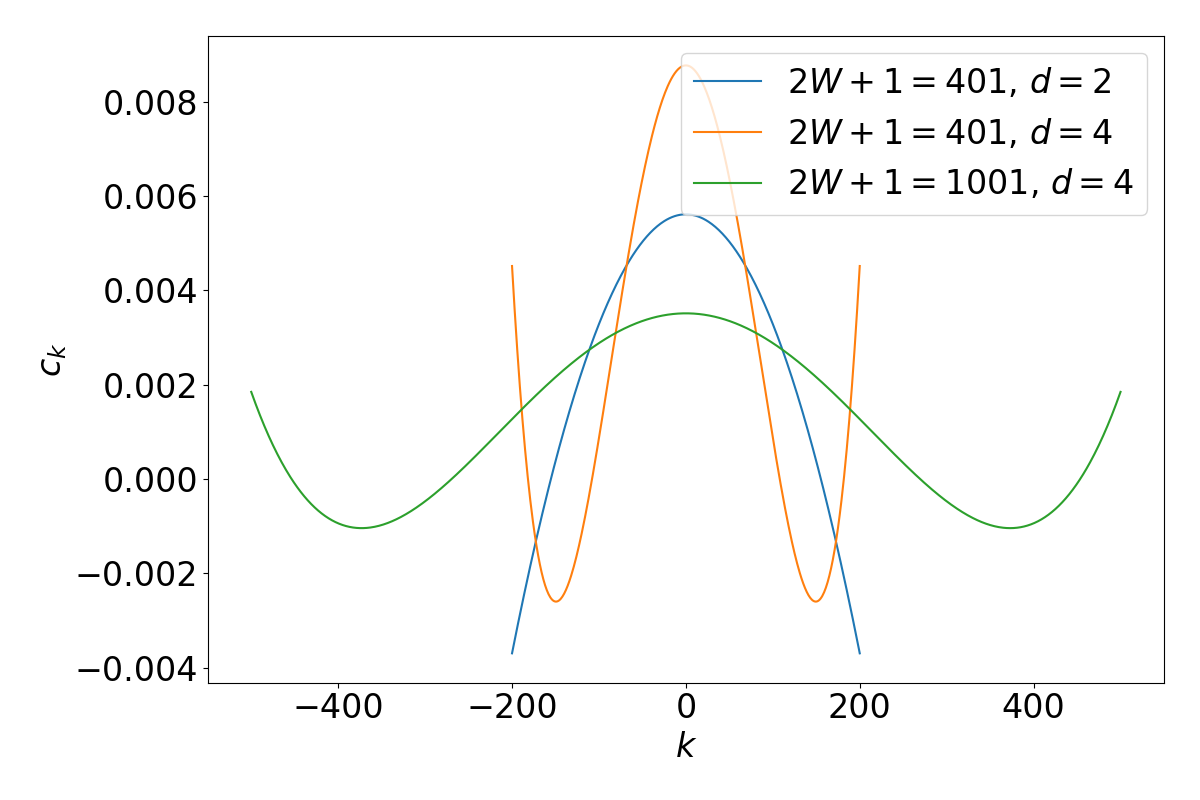}
	\caption
	{Left panel shows the power excess of an axion signal at 1.096
          GHz before and after applying an SG filter ($2W+1=401$,
          $d=4$). The power excess is in arbitrary units (a.u.) and
          the peak value of the lineshape was set to one. Right panel
          shows the SG filter coefficients $c_{k}$ for different $W$
          and $d$. For any $\abs{k}>W$, $c_{k}$ is not defined and set
          to zero in any relations using it.}        
	\label{fig_single_spectrum_signal}
\end{figure}

We will first proceed to discuss how to obtain the value of
$\epsilon_{\rm  sig}$. Using an SG filter as a background estimate, it
is possible to construct a signal following
equation~(\ref{EQ:LINESHAPE}) on a flat background that does not
include any noise fluctuations. The input signal in units of arbitrary
power excess will be distributed into $n_{c} = \Delta\nu_{a}/\Delta\nu$
frequency bins that have a resolution bandwidth $\Delta\nu$, assuming
that the first bin's frequency is equal to $\nu_{a}$. The input signal
power in the $k$th bin is
\begin{equation}
	\delta_{k} = \delta_{a}L_{k} = \delta_{a}\int_{ \nu_{a} + \left(k - \frac{1}{2}\right) \Delta\nu }^{ \nu_{a} + \left(k + \frac{1}{2}\right) \Delta\nu } f(\nu) \dd\nu ,
\end{equation}
where $\delta_{a}$ is a multiplicative constant that represents
$P_a^{a\gamma\gamma}$ and $L_{k}$ are the fractions of axion signal
power stored into the $k$th bin. While our SG filter is designed to
describe the background only, it will be applied to the set of
$\delta_{k}$ (the red line in the left panel of
Fig.~\ref{fig_single_spectrum_signal}) to parametrize the spectrum
containing the signal as well as the flat background. As also seen in
the left panel of Fig.~\ref{fig_single_spectrum_signal}, the smoothing
from the filter, shown as orange dashed lines, does not fit the
background properly around the signal region when one is present. The
exact response of the SG filter is
\begin{equation}
	\hat{\delta}_{k} = \sum_{k^{\prime} = -W}^{W} c_{k^{\prime}}\delta_{k+k^{\prime}},
\end{equation}
with $\sum_{k}c_{k}=1$. $\hat{\delta}_{k}$ is the SG filter's estimate
of the background for the cases considered in this work. What is
considered to be the signal power after background subtraction will be
$\delta^{\prime}_{k} = \delta_{k} - \hat{\delta}_{k}$ (the blue line
in the left panel of Fig.~\ref{fig_single_spectrum_signal}).
The coefficients $c_{k}$ only depend on the SG filter parameters,
window length $2W+1$ and polynomial order $d$. As $\sum_{k}c_{k}=1$,
an estimation done with the SG filter acts like a weighted
average. The same $2W+1$ values are applied to the coefficients when
obtaining each separate $\hat{\delta}_{k}$ with different $k$. A few
examples of SG filter coefficients depending on the parameters can be
seen in the right panel of Fig.~\ref{fig_single_spectrum_signal}. The
coefficients can be obtained from its definition in the original
paper~\cite{SGFILTER}, and it can be fit using a $d$th-order
polynomial (for even $d$). It is also easy to obtain $c_{k}$ in SG
filter packages in programming languages~\cite{PYTHON}.

$\delta_{k}$ and $\delta_{k}^{\prime}$ are combined according to the
analysis procedure~\cite{HAYSTAC}. Following this all further
combination processes use inverse-variance weighting. For any stage of
combination, this means that a combined power
$\delta^{\mathrm{combined}}$ and its error
$\sigma^{\mathrm{combined}}$ will follow
\begin{equation} \label{eq_combination}
	\begin{aligned}
		\delta^{\mathrm{combined}} &= \frac{\sum_{k} w_{k}\delta_{k}^{\mathrm{before}}}{\sum_{k}w_{k}} \\
		\sigma^{\mathrm{combined}} &= \frac{\sqrt{\sum_{k} w_{k}^{2}(\sigma_{k}^{\mathrm{before}})^{2}}}{\sum_{k}w_{k}}		
	\end{aligned}
\end{equation}
for all relevant powers $\delta_{k}^{\mathrm{before}}$ and their errors
$\sigma_{k}^{\mathrm{before}}$ used for the combination. In the
presence of $\sigma_{k}^{\mathrm{before}}$, the weights
$w_{k} = (\sigma_{k}^{\mathrm{before}})^{-2}$,
except during the coadding process to obtain the grand spectrum.
In this case $w_{k} = (\sigma_{k}^{\mathrm{before}})^{-2}L_{k}$.

\begin{figure}[h]
	\centering
	\includegraphics[width=0.48\columnwidth]{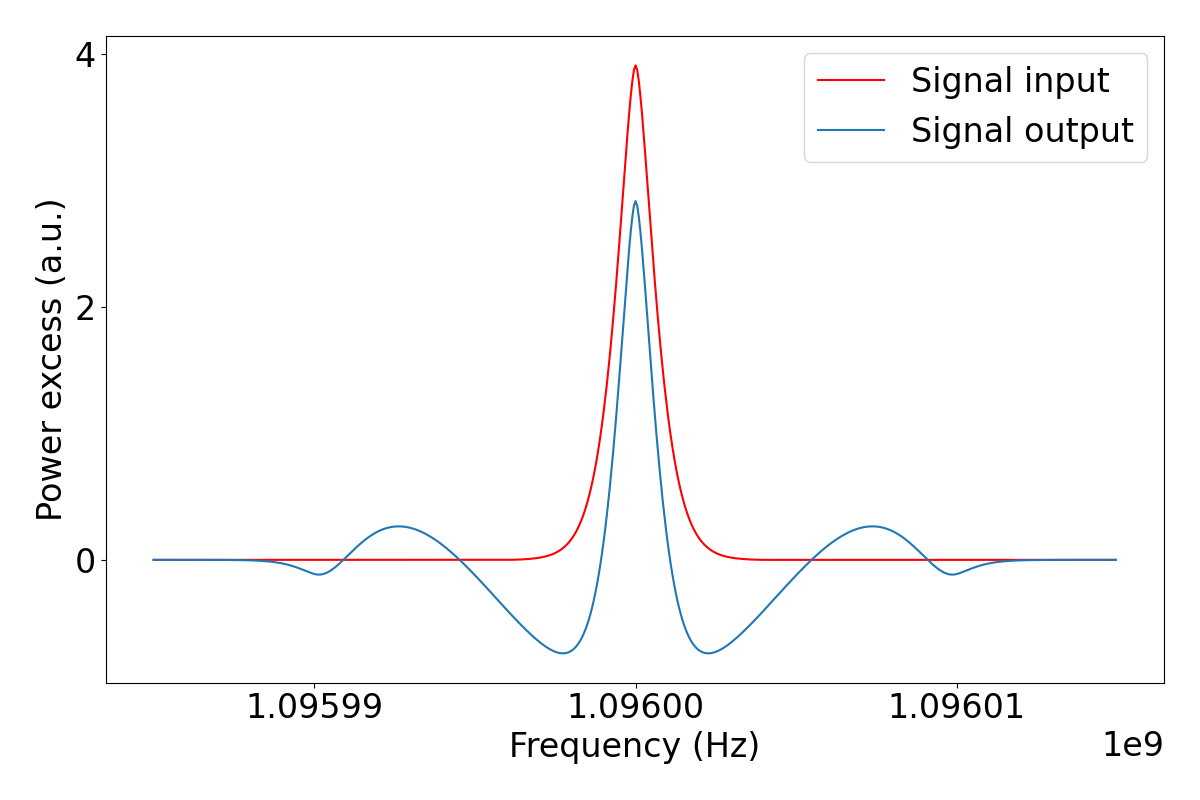}  \quad
	\includegraphics[width=0.48\columnwidth]{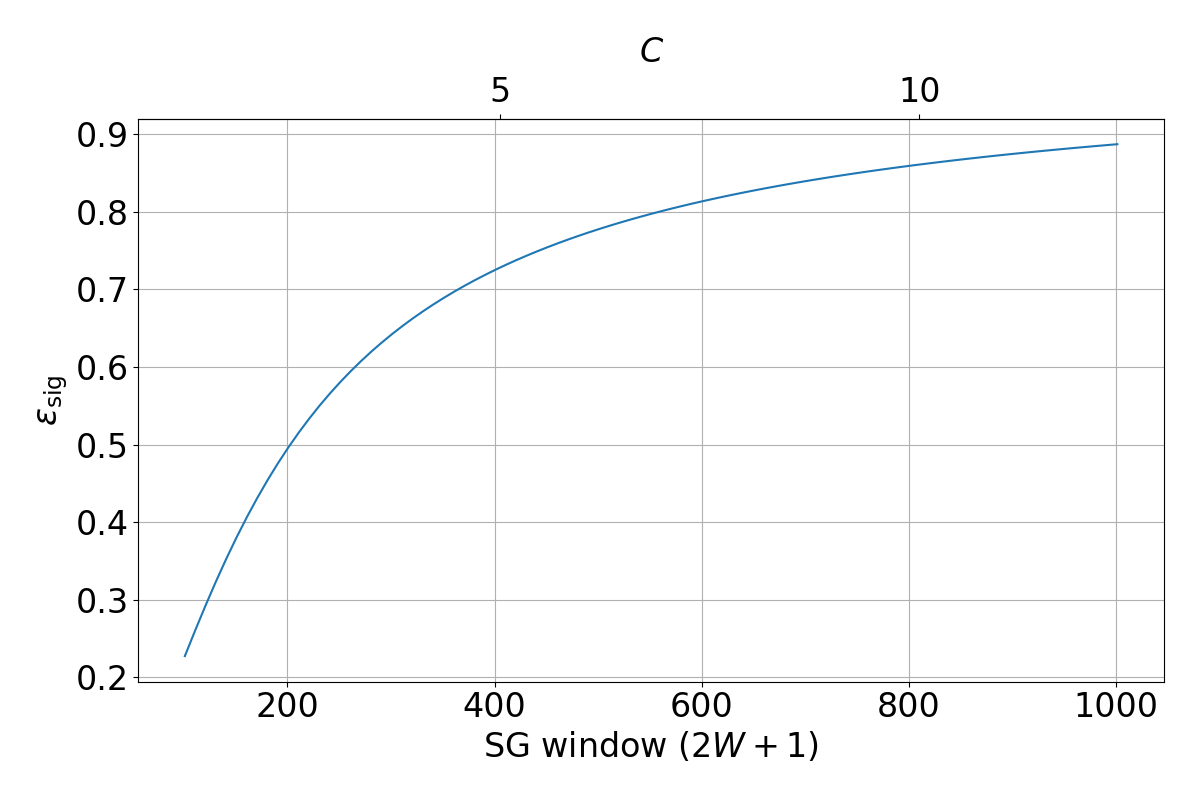}
	\caption
	{Left panel shows the input and output coadded axion signal power at 1.096 GHz.
          $\epsilon_{\rm  sig}$ is calculated when $\Delta\nu_{a} =
          4050$ Hz and $\Delta\nu=50$ Hz at the bin coinciding with
          the axion frequency $\nu_{a} = 1.096$ GHz, which is where
          the spectrum shows maximum power excess. Right panel shows
          the calculated $ \epsilon_{\rm sig}$ value at the axion
          frequency. The SG filter window is also represented with a
          normalized value $C = (2W+1) / (\Delta\nu_{a}/\Delta\nu)$.}        
	\label{fig_single_spectrum_coadded}        
\end{figure}

Now we will obtain the SNR ratio through the axion signal power ratio:
$\epsilon_{\rm  sig} = \delta_{\rm output}/\delta_{\rm input}$. As the
SG filter response of each $\delta_{k}$ is up to the
multiplicative constant $\delta_{a}$, its effects cancel out
in the value of $ \epsilon_{\rm  sig}$. Therefore the value of
$\epsilon_{\rm  sig}$ will not change when signal power is added in a
way that changes $\delta_{a}$ by the same ratio for both input and
output, such as vertical combination. This was confirmed to be in
agreement with large-statistic simulations which do include vertical
combination and will be further discussed in
Sec. \ref{sec_mc_comparison}. Rebinning, or equivalently RBW
reduction, however, does affect the results and will be discussed in
Sec. \ref{sec_rebinning}. If this process is not used, the only
leftover contribution to the analysis process is coadding bins. We can
now use equation~(\ref{eq_combination}) with
$\delta_{k}^{\rm before} = \delta_{k}$ or
$\delta^{\prime}_{k}$ and $w_{k} = L_{k}$. For the particular bin that
coincides with the axion frequency, the signal power after coadding
$n_{c}$ bins is
\begin{equation}
	\delta_{\rm input} =  \frac{\sum_{k=0}^{n_{c}-1}L_{k}\delta_{k}}{\sum_{k=0}^{n_{c}-1}L_{k}} = \frac{\delta_{a}\sum_{k=0}^{n_{c}-1}L_{k}^{2}}{\sum_{k=0}^{n_{c}-1}L_{k}}
\end{equation}
and
\begin{equation}
	\delta_{\rm output} = \frac{\sum_{k=0}^{n_{c}-1}L_{k}\delta_{k}^{\prime}}{\sum_{k=0}^{n_{c}-1}L_{k}} =  \frac{\delta_{a}\left(\sum_{k=0}^{n_{c}-1}L_{k}^{2} - \sum_{i=0}^{n_{c}-1}\sum_{j=0}^{n_{c}-1}c_{i-j}L_{i}L_{j}\right)}{\sum_{k=0}^{n_{c}-1}L_{k}}
\end{equation}
resulting in
\begin{equation}
	 \epsilon_{\rm  sig} = \frac{\mathrm{SNR}_{\rm output}}{\mathrm{SNR}_{\rm input}} \simeq \frac{\delta_{\rm output}}{\delta_{\rm input}} = 1 - \frac{ \sum_{i=0}^{n_{c}-1}\sum_{j=0}^{n_{c}-1}c_{i-j}L_{i}L_{j}} {\sum_{k=0}^{n_{c}-1}L_{k}^{2}}
\end{equation}
where if $\abs{i-j} > W$, $c_{i-j} = 0$. As $ \epsilon_{\rm  sig}$
does not take any noise fluctuations into account it does not carry
any error. Hence, this can be used as an exact value instead of an
estimate. The value of $ \epsilon_{\rm  sig}$ depending on $W$ is
shown in the right panel of Fig.~\ref{fig_single_spectrum_coadded}.

The remaining factor in $\epsilon_{\rm SNR}$, $\xi$, will be obtained
next. For this calculation there will be no axion signal, and only the
background is considered. This time the inputs will be multiple
independent random variables $x_{k}$ that follow a Gaussian
distribution
$\mathcal{N}\left(\bar{x}_{k}, \sigma_{x_{k}}^{2}\right)$.
The normalized power is
\begin{equation}
	X_{k} = \frac{x_{k}}{\sigma_{x_{k}}}
\end{equation}
making $\sigma_{X_{k}}$ one. The estimated background using the SG
filter for $X_{k}$ is
\begin{equation} \label{eq_sg_noise}
	\hat{X}_{k} = \sum_{k^{\prime}=-W}^{W} c_{k^{\prime}}X_{k + k^{\prime}} .
\end{equation}
The power excess is $X^{\prime}_{k} = X_{k} - \hat{X}_{k}$. Ideally
the estimate $\hat{X}_{k}$ is equal to
$\bar{X}_{k} = \bar{x}_{k}/\sigma_{x_{k}}$ as $X_{k}$ follows an
$\mathcal{N}\left(\bar{X}_{k}, \sigma_{X_{k}}^{2}\right)$
distribution.
It was observed~\cite{12TB-PRD} and is known that the standard
deviation of $X^{\prime}_{k}$ depends on $W$ and is reduced by 
\begin{equation}
	\xi_{\rm SG} = 1 - \frac{b}{2W + 1},
\end{equation}
where $b=9/8$ for $d=2$ and $b=225/128$ for $d=4$ if $W$ is large
enough~\cite{SGFILTER-PROPERTIES}.
Therefore $X^{\prime}_{k}$ must be further divided by $\xi_{\rm SG}$
to make it follow an $\mathcal{N}\left(0,\sigma_{X^{\prime}_{k}}^{2}\right)$ distribution with
$\sigma_{X^{\prime}_{k}}=1$. The normalized power excess for SG filter
estimated data is
\begin{equation} \label{eq_noise_baseline}
	X^{\prime}_{k} = \frac{1}{\xi_{\rm SG}} \left(X_{k} - \sum_{k^{\prime}=-W}^{W} c_{k^{\prime}}X_{k + k^{\prime}} \right) .
\end{equation}

The use of normalized power excess to obtain $\xi$ is valid as it is
independent of the background~\cite{HAYSTAC, JHEP} and so are the
coefficients $c_{k}$. The exception to this would be when background
subtraction is done poorly or is biased, which will be explored in
Sec.~\ref{sec_mc_comparison}.
Here we will continue with the assumption that $\hat{X}_{k}$ is an
unbiased estimate of $\bar{X}_{k}$. Similar to the discussion from
before, a vertical combination combines different raw spectra, so the
random variables that are combined must be independent, and therefore
it does not affect the distribution of $X_{k}$ nor $X^{\prime}_{k}$.

\begin{figure}[h]
	\centering
	\includegraphics[width=0.6\columnwidth]{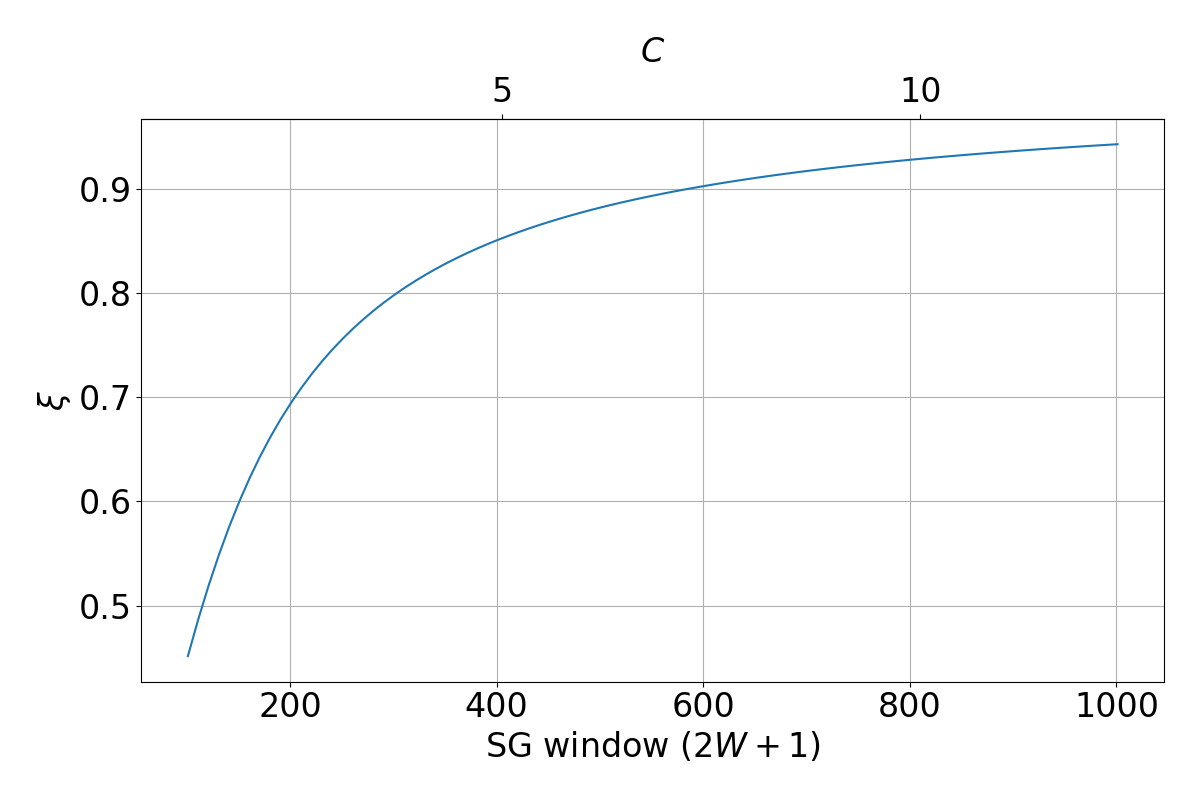}
	\caption
	{The calculated value of $\xi$ using an SG filter with
          $d=4$. The $L_{k}$ used for coadding is the same as those
          used to calculate $\epsilon_{\rm sig}$ above. The SG filter
          window is also represented with a normalized value
          $C =(2W+1) / (\Delta\nu_{a}/\Delta\nu)$.}
	\label{fig_sigma_guess}
\end{figure}

Now equation~(\ref{eq_combination}) is used again with
$\delta_{k}^{\rm before} = X_{k}$ or
$X^\prime_{k}$ and $w_{k} = L_{k}$ with
$\sigma_{k}^{\rm before} = \sigma_{X_{k}}$ or
$\sigma_{X^{\prime}_{k}}$. The normalized input noise at the axion
frequency corresponding to $k=0$ after coadding is
\begin{equation}
	X_{\rm coadded} = \frac{\sum_{k=0}^{n_{c}-1}L_{k}X_{k}}{\sqrt{\sum_{k=0}^{n_{c}-1}L_{k}^{2}}} ,
\end{equation}
which has a standard deviation of one as expected from $n_{c}$
independent random variables.
Coadding the set of $X^{\prime}_{k}$ we have
\begin{equation}
	X_{\rm coadded}^{\prime} = \frac{\sum_{k=0}^{n_{c}-1}L_{k}X^{\prime}_{k}}{\sqrt{\sum_{k=0}^{n_{c}-1} L_{k}^{2}}} = \frac{1}{\xi_{\rm SG}} \frac{\sum_{i=-W}^{W+n_{c}-1}C_{i}X_{i}}{\sqrt{\sum_{k=0}^{n_{c}-1}L_{k}^{2}}} .
\end{equation}
$X^{\prime}_{k}$ are no longer independent to each other since each
can include overlapping terms that can range from $X_{-W}$ to
$X_{W+n_{c}-1}$ according to
equation~(\ref{eq_noise_baseline}). Therefore this equation is used to
decouple $X_{\rm coadded}^{\prime}$ into independent random variables
$X_{i}$. The coefficient value in front of $X_{i}$ is
\begin{equation}
	C_{i} = L_{i} - \sum_{k=0}^{n_{c}-1} c_{i-k}L_{k}
\end{equation}
with the constraints $L_{i} = 0$ for $i < 0$ or $i \ge n_{c}$,
$c_{k} = 0$ for $k < -W$ or $k > W$.
The standard deviation of $X_{\rm coadded}^{\prime}$, which
corresponds to the frequency independent scale factor $\xi$, is
\begin{equation} \label{eq_sigma_guess}
	\xi = \frac{1}{\xi_{\rm SG}} \frac {\sqrt{\sum_{i=-W}^{W+n_{c}-1}C_{i}^{2}}}{\sqrt{\sum_{k=0}^{n_{c}-1}L_{k}^{2}}} .
\end{equation}
The value of $\xi$ for an SG filter with varying $W$ can be seen in
Fig.~\ref{fig_sigma_guess}.

It can be observed that both $ \epsilon_{\rm  sig}$ and $\xi$ become
closer to unity as $W$ increases. This can be understood intuitively
since the SG filter is an estimate for a background in the context of
this raw spectrum. As $W$ becomes larger, the SG filter will tend to
``ignore" the signal and therefore retain more of its power, without
being too biased. On the other hand, for noise fluctuations when more
points are taken into consideration, the bin-to-bin
correlations are smaller as $W$ increases since the sum of all
$c_{k}$ is one, meaning that the contribution of each bin is
smaller, as seen from the right panel of
Fig.~\ref{fig_single_spectrum_signal}. This also tends to average out to
$\bar{X}_{k}$.

\section{Considering the effects of rebinning spectra} \label{sec_rebinning}
By merging $n_{m}$ bins and increasing the RBW value to
$\Delta\nu^{\rm merged} = n_{m}\Delta\nu$, the power spectrum
undergoes a process called rebinning. This process combines the
$n_{m}$ bins according to equation~(\ref{eq_combination}).
For the calculation of $\epsilon_{\rm  sig}$, the signal powers after
rebinning with $w_{k}=1$ and $\delta^{\rm before}_{k}=\delta_{k}$ are
\begin{equation}
	\begin{aligned}
		\delta_{k}^{\rm merged} = \frac{\sum_{i}\delta_{i}}{n_{m}}, \\
		\delta_{k}^{\prime\rm merged} = \frac{\sum_{i}\delta^{\prime}_{i}}{n_{m}},
	\end{aligned}
\end{equation}
which are the averages of the signal powers over $n_{m}$ bins.

A coadded bin after rebinning will coadd $n_{c}/n_{m}$ bins, with the
weights used for coadding $w_{k} = L_{k}^{\rm merged}$ following
\begin{equation}
	L_{k}^{\rm merged} = \int_{ \nu_{a} + \left(k - \frac{1}{2}\right) \Delta\nu^{\rm merged} }^{ \nu_{a} + \left(k + \frac{1}{2}\right) \Delta\nu^{\rm merged} } f(\nu) \dd\nu .
\end{equation}
Obtaining $ \epsilon_{\rm  sig}$ is straightforward in the sense that
the rebinning is applied to each $\delta_{k}$, then the rebinned bins
are coadded.
\begin{equation}
	\begin{aligned}
		\delta^{\rm merged}_{\rm input} &= \frac{\sum_{k=0}^{n_{c}/n_{m} - 1} L_{k}^{\rm merged}\delta_{k}^{\rm merged}}{\sum_{k=0}^{n_{c}/n_{m} - 1} L_{k}^{\rm merged}} \\
		\delta^{\rm merged}_{\rm output} &= \frac{\sum_{k=0}^{n_{c}/n_{m} - 1} L_{k}^{\rm merged}\delta_{k}^{\prime\rm merged}}{\sum_{k=0}^{n_{c}/n_{m} - 1} L_{k}^{\rm merged}} 
	\end{aligned}
\end{equation}
and $\epsilon^{\rm merged}_{\rm  sig} = \delta^{\rm merged}_{\rm output} / \delta^{\rm merged}_{\rm input}$.
For both signal input and output the resulting SNR values have been
shown to decrease, as seen in ref.~\cite{12TB-PRL}.
While this is a drawback of rebinning, the number of rescan candidates
is also reduced, which is an advantage that could be considered as a
tradeoff~\cite{12TB-PRL, 12TB-PRD}.

Now the rebinned counterpart for noise, $X_{k}^{\rm merged}$ and
$X_{k}^{\prime\mathrm{merged}}$, will yield a modified
$\xi^{\rm merged}$. Using equation~(\ref{eq_sigma_guess}) and the fact $X_{k}$
and $X_{k}^{\prime}$ are combined in a linear fashion, the only part
of the equation that needs modification is $L_{k}$, which must be
expressed in terms of $L_{k}^{\rm merged}$. For any $n_{m}$ bins
merged, $X_{k}^{\rm merged}$ and $\sigma_{X_{k}^{\rm merged}}$ are as
follows
\begin{equation} \label{eq_merged_noise}
	\begin{aligned}
		X_{k}^{\rm merged} &= \frac{\sum_{i}X_{i}}{n_{m}} ,\\
		\sigma_{X_{k}^{\rm merged}} &= \frac{1}{\sqrt{n_{m}}},
	\end{aligned}
\end{equation}
where the total number of $i$ is $n_m$. The same applies for
$X_{k}^{\prime\rm merged}$. 
All $n_{m}$ merged bins will now be coadded with the same
$L_{k}^{\rm merged}$ value instead of the $n_{m}$ different $L_{k}$ values
(e.g., $L_{0}$, $L_{1}$, $\cdots$, $L_{n_{m}-1}$) from before. 
The coadded and normalized values are
\begin{equation} \label{eq_merged_coadded}
  \begin{aligned}
    \mathcal{X}_{\rm coadded} &= \frac{X_{\rm coadded}}{\sigma_{X_{\rm coadded}}}=\frac{\sqrt{n_m}\sum_{k=0}^{n_{c}/n_{m}-1}L^{\rm merged}_{k} X^{\rm merged}_{k}}{\sqrt{\sum_{k=0}^{n_{c}/n_{m}-1}\left(L^{\rm merged}_{k}\right)^{2}}} = \frac{\sum_{k=0}^{n_{c}-1}L^{\rm merged}_{\lfloor{k/n_{m}}\rfloor}X_k}{\sqrt{\sum_{k=0}^{n_{c}-1} \left(L^{\rm merged}_{\lfloor{k/n_{m}}\rfloor}\right)^{2}}}, \\
    \mathcal{X}^{\prime}_{\rm coadded} &= \frac{X^{\prime}_{\rm coadded}}{\sigma_{X^{\prime}_{\rm coadded}}}=\frac{\sum_{k=0}^{n_{c}-1}L^{\rm merged}_{\lfloor{k/n_{m}}\rfloor}X^{\prime}_k}{\sqrt{\sum_{k=0}^{n_{c}-1}\left(L^{\rm merged}_{\lfloor{k/n_{m}}\rfloor}\right)^{2}}} = \frac{1}{\xi_{\rm SG}}\frac{\sum_{i=-W}^{W+n_{c}-1}C^{\rm merged}_{i} X_i}{\sqrt{\sum_{k=0}^{n_{c}-1} \left(L^{\rm merged}_{\lfloor{k/n_{m}}\rfloor}\right)^{2}}}, 
  \end{aligned}
\end{equation}
where the floor function $\lfloor{x}\rfloor$ returns the largest
integer smaller or equal to $x$.
$\mathcal{X}_{\rm coadded}$ has a standard deviation of one as expected. 
For $\mathcal{X}^{\prime}_{\rm coadded}$, the standard deviation, or
equivalently, the frequency independent scale factor after rebinning
$\xi^{\rm merged}$ is obtained as
\begin{equation} \label{eq_sigma_guess_merged}
	\xi^{\rm merged} = \frac{1}{\xi_{\rm SG}} \frac {\sqrt{\sum_{i=-W}^{W+n_{c}-1}(C_{i}^{\rm merged})^{2}}}{\sqrt{\sum_{k=0}^{n_{c}-1}(L^{\rm merged}_{\lfloor{k/n_{m}}\rfloor})^{2}}} 
\end{equation}
and the relevant results are shown in table~\ref{table_mc_comp}.
$C_{i}^{\rm merged}$ is also different from
$C_{i}$ since it also includes  $L^{\rm merged}_{k}$ as
\begin{equation} \label{eq_merged_coadded_coeff}
  C_{i}^{\rm merged} = L^{\rm merged}_{\lfloor{i/n_{m}}\rfloor} - \sum_{k=0}^{n_{c}-1} c_{i-k}L^{\rm merged}_{\lfloor{k/n_{m}}\rfloor} .
\end{equation}
As shown in equation~(\ref{eq_merged_coadded}), the rebinned set of
$X^{\prime\rm merged}_{k}$ in equation~(\ref{eq_merged_coadded}) is
separable into $X^{\prime}_{k}$ using equation~(\ref{eq_merged_noise})
and ultimately $X_{k}$ via equation~(\ref{eq_noise_baseline}). By
unpacking the merged variables fully into independent random
variables, the calculation of $\xi^{\rm merged}$ in
equation~(\ref{eq_sigma_guess_merged}) becomes as straightforward as
$\xi$ in equation~(\ref{eq_sigma_guess}).

Putting the effects of rebinning from both sides together, the SNR
efficiency $\epsilon_{\rm SNR}^{\rm merged} = \epsilon^{\rm merged}_{\rm  sig}  / \xi^{\rm merged}$ is smaller than
$\epsilon_{\rm SNR}$, which is a separate effect from the
aforementioned SNR decrease reported in ref.~\cite{12TB-PRL}.
This is later shown in table~\ref{table_mc_comp} ($\epsilon_{\rm SNR}$
and $\epsilon_{\rm SNR}^{\rm merged}$) and Fig.~\ref{fig_eff_comp}
(blue and orange lines), although the difference is smaller for large $W$. While
$\epsilon^{\rm merged}_{\rm  sig} < \epsilon_{\rm  sig}$ and $\xi^{\rm merged} < \xi$,
the difference between $\epsilon_{\rm  sig}$ and its rebinned
counterpart is larger than the change in $\xi$, leading to a decrease
in SNR efficiency after rebinning.

\section{Comparison with large-statistic simulations} \label{sec_mc_comparison}
\begin{figure}[h]
	\centering
	\includegraphics[width=0.6\columnwidth]{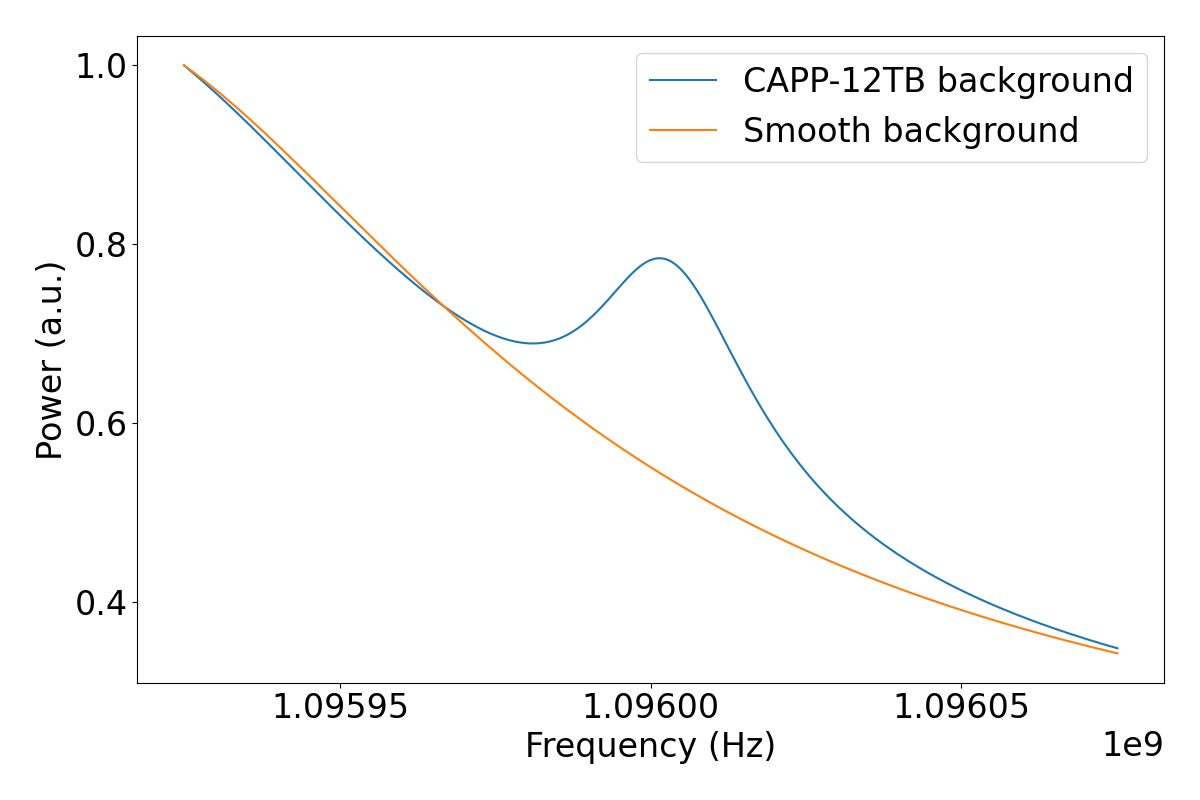}
	\caption
	{A comparison between the backgrounds used for large-statistic simulations.}
	\label{fig_bg_comp}
\end{figure}

So far the value of $\epsilon_{\rm SNR}$ has been obtained given $\nu_{a}$,
$\Delta\nu_{a}$, and $\Delta\nu$. In order to see if this agrees with
previously used large-statistic simulations, an axion signal was
inserted by software using realistic CAPP-12TB background data. The
input CAPP-12TB background shapes use the function and its parameters
that were obtained from the $\chi^{2}$ fit in the original data
analysis~\cite{12TB-PRL}. Another background was constructed for
comparison as well, which used the same parameters except for one that
described the cavity bandwidth, which was multiplied by 20. This
ensured that the structure of the cavity was not seen in this
background, and instead was a ``smooth" one. Both backgrounds are
compared in Fig.~\ref{fig_bg_comp}. Each simulation was done
10000 times with pseudo-random-generated noise to obtain a sufficient
amount of statistics, with an axion signal injected near 1.096 GHz. We
will denote the results for the two types of backgrounds as ``12TB"
and ``smooth," respectively, e.g., $\epsilon_{\rm SNR}^{\rm 12TB}$ or
$\xi^{\rm smooth}$. Since the simulations follow the CAPP-12TB
analysis parameters, there is rebinning. Hence the relevant values
should be compared with $\epsilon_{\rm SNR}^{\rm merged}$ and
$\xi^{\rm merged}$.

\begin{figure}[h]
	\centering
	\includegraphics[width=0.6\columnwidth]{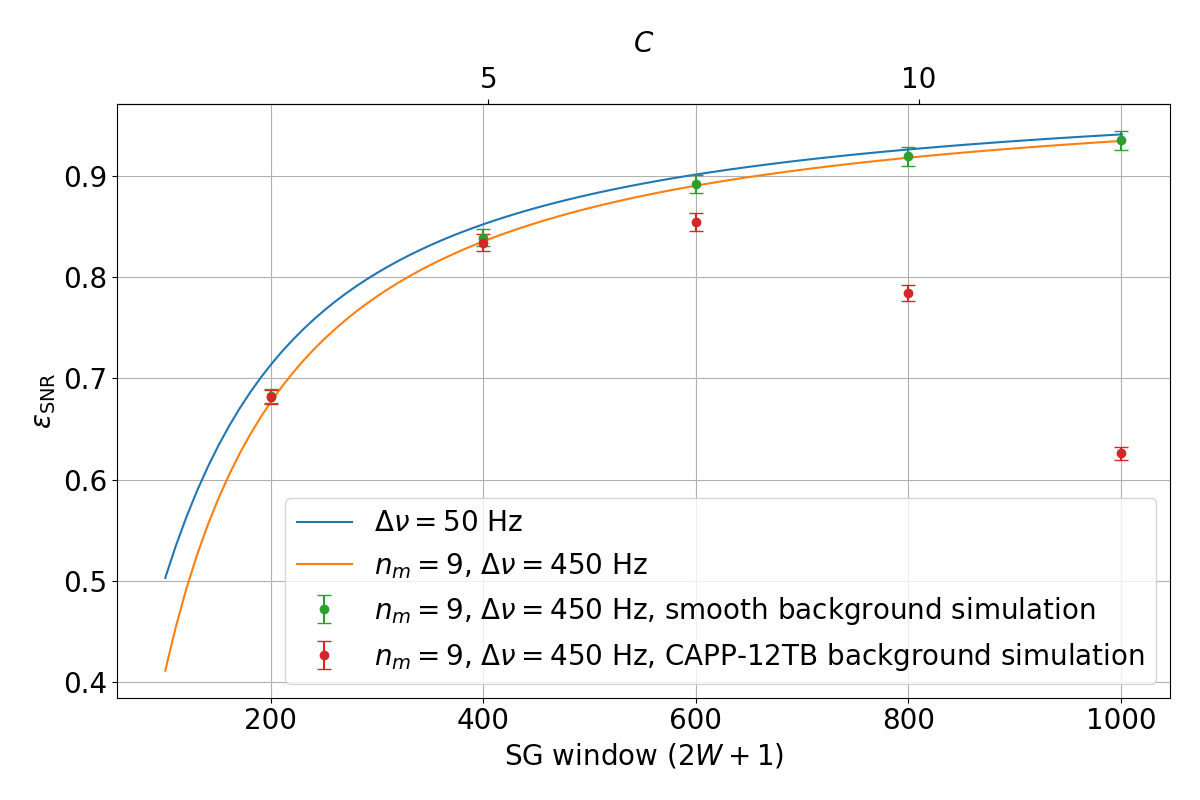}
	\caption
	{A comparison of the calculated $\epsilon_{\rm SNR}$
          values and large-statistic simulation $\epsilon_{\rm SNR}$
          values for an SG filtered axion signal at $\nu_{a} = 1.096$
          GHz with $\Delta\nu_{a} = 4050$ Hz. The blue line represents
          $\epsilon_{\rm SNR}$ after coadding at $\Delta\nu=50$ Hz and
          the orange line represents $\epsilon_{\rm SNR}$ after
          rebinning with $n_{m}=9$ bins merged ($\Delta\nu^{\rm merged} = 450$ Hz)
          and then coadding. The green and red
          dots are results from simulations with different backgrounds
          and comes with 1\% error (10000 averages). The SG filter
          window is also represented by a normalized value
          $C = (2W+1) / (\Delta\nu_{a}/\Delta\nu)$.}        
	\label{fig_eff_comp}
\end{figure}

Compared to the time and disk space needed for large-statistic
simulations, obtaining the values analytically for the same parameters
is practically interactive, taking less than a minute while using only
computer memory. The results can be compared in
table~\ref{table_mc_comp} and Fig.~\ref{fig_eff_comp}.
\begin{table} [h]
	\centering
	\begin{tabular}{ | c | c | c | c | c | c | c | c | c | } 
		\hline
		& \multicolumn{2}{| c |}{Analytical} & \multicolumn{2}{| c |}{Simulation} & \multicolumn{2}{| c |}{Analytical} & \multicolumn{2}{| c |}{Simulation} \\ \hline
		$2W+1$ & $\xi$ & $\xi^{\rm merged}$ & $\xi^{\rm smooth}$ & $\xi^{\rm 12TB}$ & $\epsilon_{\rm SNR}$ & $\epsilon_{\rm SNR}^{\rm merged}$ & $\epsilon_{\rm SNR}^{\rm smooth}$ & $\epsilon_{\rm SNR}^{\rm 12TB}$ \\ \hline
		201 & 0.6948 & 0.6567 & 0.6540 & 0.6537 & 0.7145 & 0.6779 & 0.6823 & 0.6819 \\ \hline  
		401 & 0.8510 & 0.8328 & 0.8314 & 0.8315 & 0.8525 & 0.8356 & 0.8388 & 0.8342 \\ \hline  
		601 & 0.9028 & 0.8912 & 0.8910 & 0.9019 & 0.9015 & 0.8905 & 0.8919 & 0.8543 \\ \hline  
		801 & 0.9286 & 0.9196 & 0.9198 & 0.9849 & 0.9262 & 0.9180 & 0.9193 & 0.7840 \\ \hline  
		1001 & 0.9429 & 0.9362 & 0.9371 & 1.1285 & 0.9410 & 0.9346 & 0.9353 & 0.6261 \\ \hline  
	\end{tabular}
	\caption{A comparison of various $\epsilon$ and $\xi$
          obtained from analytic calculation or simulations. The
          polynomial order for all SG filters was $d=4$. As the
          simulation results include rebinning, $\xi^{\rm smooth}$ and
          $\xi^{\rm 12TB}$ should be compared with $\xi^{\rm merged}$,
          while  $\epsilon_{\rm SNR}^{\rm smooth}$ and
          $\epsilon_{\rm SNR}^{\rm 12TB}$ should be compared with
          $\epsilon_{\rm SNR}^{\rm merged}$.
          Efficiency results on the simulation side include a 1\%
          error.
          The results do not consider the additional 5\% loss in SNR
          that occurs due to rebinning, i.e.,
          $\mathrm{SNR}_{\mathrm{input}}$ and          
          $\mathrm{SNR}_{\mathrm{output}}$ are both at the
          $\Delta\nu = 450$ Hz level when calculating
          $\epsilon_{\rm SNR}^{\rm merged}$, $\epsilon_{\rm SNR}^{\rm smooth}$,
          and $\epsilon_{\rm SNR}^{\rm 12TB}$.}        
	\label{table_mc_comp}
\end{table}

As seen from the simulation results, a smooth background matches the
rebinned analytical calculations well, but the 12TB background does
not. This is expected from ref.~\cite{SGFILTER-PROPERTIES}, i.e., as
$W$ increases the background estimation begins to show bias for the
12TB background shown in Fig.~\ref{fig_bg_comp}, similar to the SG
filter's tendency to ``ignore" the axion signal, as mentioned
earlier. Looking at $\xi^{\rm 12TB}$, $\xi^{\rm smooth}$, and
$\xi^{\rm merged}$ in table~\ref{table_mc_comp}, this bias propagates
to the grand spectrum and becomes evident from $2W+1=801$.
Furthermore, the bias even shows a positive correlation when
$2W+1=1001$, which has also been noted in~\cite{HAYSTAC}. In terms of
$\epsilon_{\rm SNR}$ in Fig.~\ref{fig_eff_comp}, which combines the
effects of $\epsilon_{\rm sig}$ and $\xi$, we can see the bias become
evident from $2W+1=601$.
\begin{figure}[h]
	\centering
	\includegraphics[width=0.6\columnwidth]{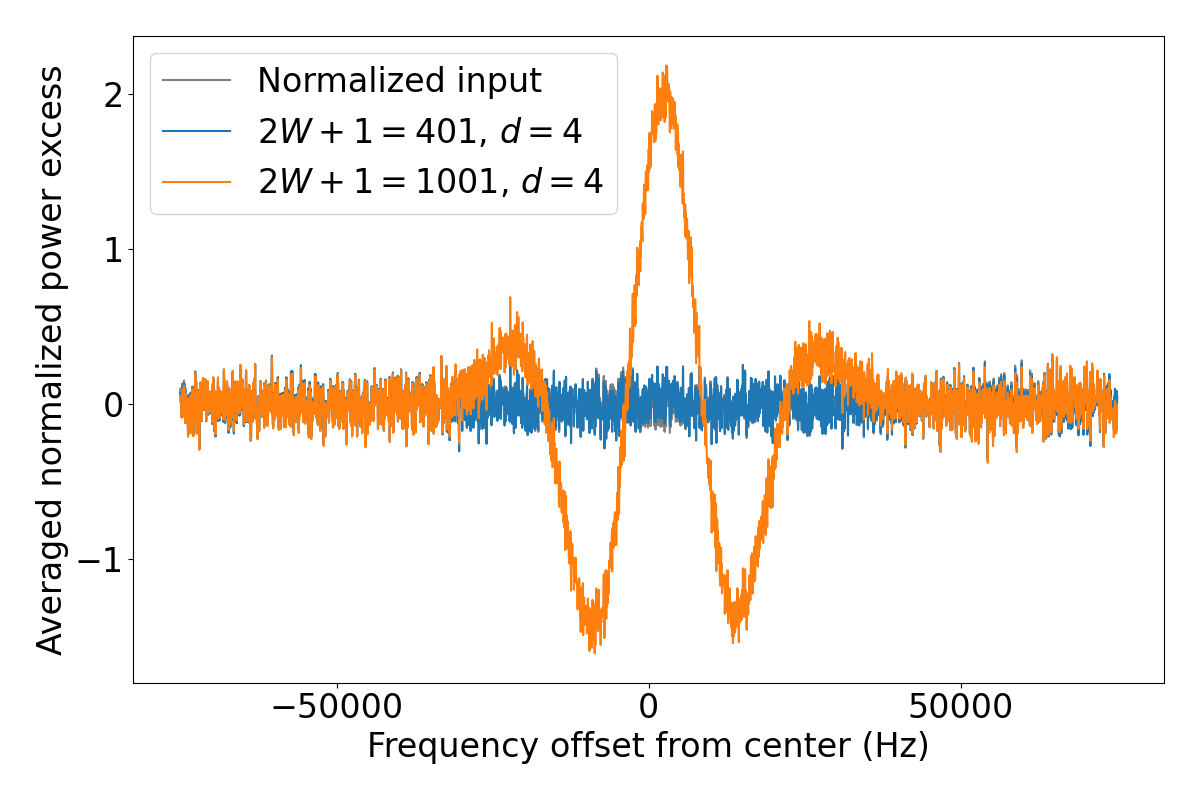}
	\caption
	{100 averaged normalized power excesses for $2W+1=401$, $d=4$ and $2W+1=1001$, $d=4$ for the CAPP-12TB background.}
	\label{fig_bg_bias}
\end{figure}

The bias shown in the normalized power excess in the CAPP-12TB
background suggests that there is a limit to using larger $W$ for SG
filters and this can affect both $\epsilon_{\rm SNR}$ and $\xi$ in an
undesirable manner.
Using the analytical calculations can work as a
time-efficient way to obtain the SNR efficiency and a predictor for
$\xi$ during data analysis. If the resulting $\xi$ from data analysis
is noticeably different compared to its predicted value, this could
indicate that there is bias in background subtraction.
As seen in Fig.~\ref{fig_bg_bias}, there is a pattern in the bias that
comes from the cavity response in the CAPP-12TB background. The bias
is not always obvious as even the $2W+1=401$ estimate shows a slight
bump near the center of the normalized power excess, but only after
one hundred different power spectra have been averaged. While this
ultimately did not dramatically affect the results in
table~\ref{table_mc_comp}, there is a certain upper limit in $W$ which
could be used before it becomes inaccurate.
This, however, does not suggest a limitation of this analytical method
but rather acts as a confidence check for background subtraction
to look out for instances such as the one shown in Fig.~\ref{fig_bg_bias}.
The verification of proper background subtraction is necessary in the
analysis procedure not only for the correct estimation of SNR but also
for obtaining unbiased power excesses.

In table~\ref{table_other_experiments}, as further validation, two
other experiments that used the SG filter for background subtraction
and provided the required parameters ($\nu_{a}$, $\Delta\nu$, $W$,
$d$, $n_{m}$) have their $\xi$ and $\epsilon_{\rm SNR}$ values
compared with the analytical method.
\begin{table} [h]
	\centering
	\begin{tabular}{ | c | c | c | c | c | c | c | c | c | c | } 
		\hline
		Experiment & $\nu_{a}$ & $\Delta\nu$ & $W$ & $d$ & $n_{m}$ & $\xi^{\dagger}$ & $\xi^{\dagger\dagger}$  & $\epsilon_{\rm SNR}^{\dagger}$ & $\epsilon_{\rm SNR}^{\dagger\dagger}$ \\ \hline
		HAYSTAC~\cite{HAYSTAC} & 5.75 GHz & 100 Hz & 500 & 4 & 10 & 0.93 & 0.919 & 0.90 & 0.909 \\ \hline  
		CAST-CAPP~\cite{CAST-CAPP} & 5.027 GHz & 50 Hz & 500 & 4 & 28 & 0.74 & 0.765 & 0.717 & 0.716 \\ \hline  
	\end{tabular}
	\caption{Comparisons of reported values from other
          experiments ($\xi^{\dagger}$ and
          $\epsilon_{\rm SNR}^{\dagger}$) with their analytically
          calculated values from this work ($\xi^{\dagger\dagger}$
          and $\epsilon_{\rm SNR}^{\dagger\dagger}$). HAYSTAC used an
          unboosted Maxwellian to obtain $\xi^{\dagger}$ and
          $\epsilon_{\rm SNR}^{\dagger}$ in their paper, which was
          also considered in our calculations.          
          The CAST-CAPP experiment initially reported $W=1000$ in
          their published paper, but the corresponding authors have
          confirmed that this was a typo and the actual parameter used
          was $2W+1=1001$ ($W=500$).}
	\label{table_other_experiments}
\end{table}

\section{Summary}
We report a fully analytical estimation of the signal to noise ratio
efficiency $\epsilon_{\rm SNR}$ in axion dark matter searches
employing an SG filter for a background estimate.
Provided the background estimation is properly done with appropriate
SG filter parameters like the blue colored data shown in Fig.~\ref{fig_bg_bias}, 
this approach provides us with the $\epsilon_{\rm SNR}$ interactively, using only an arbitrary axion mass 
and the relevant shape information, without relying on large-statistic simulation data, reflecting the
background information and the expected axion signal power obtained from a real experiment. 
By comparing our $\xi$ or $\xi^{\rm merged}$ with the width of the
normalized grand power spectrum from the experimental data,
one can also quickly cross check the reasonableness of the error estimation as
well as contamination of non-Gaussian components in the experimental data.
Axion haloscope searches have been observing the
coincidence that the frequency independent scale factor $\xi$ is
approximately consistent with $\epsilon_{\rm SNR}$. This was
confirmed analytically in this work, when the window length of the SG
filter is reasonably wide enough, i.e., at least 5 times the signal
window.

\acknowledgments
This work is supported by the Institute for Basic Science (IBS) under
Project Code No. IBS-R017-D1-2023-a00.


\begin{thebibliography}{99}

\bibitem{PLANCK}
  P. A. R. Ade {\it et al.} (Planck Collaboration), Astron. Astrophys. \textbf{594} (2016) A13. 

\bibitem{CDM-EVIDENCE}
  V. Rubin and W. K. Ford Jr., ApJ \textbf{159} (1970) 379;
  Douglas Clowe {\it et al.}, ApJ \textbf{648} (2006) L109.
  
\bibitem{AXION}
  S. Weinberg, Phys. Rev. Lett. \textbf{40} (1978) 223;
  F. Wilczek, Phys. Rev. Lett. \textbf{40} (1978) 279.

\bibitem{PQ}
  R. D. Peccei and H. R. Quinn, Phys. Rev. Lett. \textbf{38} (1977) 1440.

\bibitem{strongCP}
  G. 't Hooft, Phys. Rev. Lett, {\bf 37} (1976) 8; 
  Phys. Rev. D {\bf 14} (1976) 3432; {\bf 18} (1978) 2199(E); 
  J. H. Smith, E. M. Purcell, and N. F. Ramsey, Phys. Rev. \textbf{108} (1957) 120;
  W. B. Dress, P. D. Miller, J. M. Pendlebury, P. Perrin, and N. F. Ramsey, Phys. Rev. D {\bf 15} (1977) 9; 
  I. S. Altarev {\it et al.}, Nucl. Phys. \textbf{A341} (1980) 269. 

\bibitem{sikivie}
  P. Sikivie, Phys. Rev. Lett. \textbf{51} (1983) 1415; Phys. Rev. D \textbf{32} (1985) 2988.

\bibitem{ADMX-DFSZ}
  N. Du {\it et al.} (ADMX Collaboration), Phys. Rev. Lett. \textbf{120} (2018) 151301;
  T. Braine {\it et al.} (ADMX Collaboration), Phys. Rev. Lett. \textbf{124} (2020) 101303;
  C. Bartram {\it et al.} (ADMX Collaboration), Phys. Rev. Lett. \textbf{127} (2021) 261803.

\bibitem{12TB-PRL}
  Andrew K. Yi {\it et al.}, Phys. Rev. Lett. \textbf{130} (2023) 071002.

\bibitem{GUT}
  Howard Georgi and S. L. Glashow, Phys. Rev. Lett. \textbf{32} (1974) 438.
  
\bibitem{scanrate}
  L. Krauss, J. Moody, F. Wilczek, and D. E. Morris, Phys. Rev. Lett. \textbf{55} (1985) 1797.

\bibitem{JINST}
  S. Ahn {\it et al.}, JINST \textbf{17} (2022) P05025.

\bibitem{DICKE}
  R. H. Dicke, Rev. Sci. Instrum. \textbf{17} (1946) 268.

\bibitem{ADMX}
  S. J. Asztalos {\it et al.}, Phys. Rev. D \textbf{64} (2001) 092003.

\bibitem{HAYSTAC}
  B. M. Brubaker, L. Zhong, S. K. Lamoreaux, K. W. Lehnert, and K. A. van Bibber, Phys. Rev D \textbf{96} (2017) 123008.

\bibitem{JHEP}
  S. Ahn, S. Lee,  J. Choi, B. R. Ko, and Y. K. Semertzidis, J. High Energ. Phys. \textbf{04} (2021) 297.
  
\bibitem{12TB-PRD}
  Andrew K. Yi {\it et al.}, Phys. Rev. D \textbf{108} (2023) L021304.
  
\bibitem{TURNER}
  M. S. Turner, Phys. Rev. D \textbf{42} (1990) 3572.
  
\bibitem{CAPP-8TB-PRL}
  S. Lee, S. Ahn, J. Choi, B. R. Ko, and Y. K. Semertzidis, Phys. Rev. Lett. \textbf{124} (2020) 101802.
   
\bibitem{SGFILTER}
  A. Savitzky and M. J. E. Golay, Anal. Chem. \textbf{36} (1964) 1627.
  
\bibitem{PYTHON}
  Python Software Foundation, \url{https://www.python.org/}.
  
\bibitem{CAST-CAPP}
  M. Buschmann {\it et al.}, Nat. Commun. \textbf{13} 1049 (2022).
  
\bibitem{SGFILTER-PROPERTIES}
  H. Zielger, Appl. Spectroscopy, \textbf{35} (1981) 1.
    
\end{thebibliography}
\end{document}